\documentclass[sigconf, screen=true]{acmart}

\usepackage[utf8]{inputenc}
\usepackage{booktabs}
\usepackage{multirow}
\usepackage{subcaption}
\usepackage{balance}
\usepackage[capitalise, noabbrev]{cleveref}
\usepackage{enumitem}
\usepackage{calc}

\setcopyright{rightsretained}

\acmConference[DIR2018]{Dutch-Belgian Information Retrieval Workshop}{Nov 2018}{Leiden, the Netherlands}
\acmYear{2018}
\copyrightyear{2018}
\acmDOI{}
\acmISBN{}

\newcommand{\upup}{$^{\blacktriangle}$}
\newcommand{\downdown}{$^{\blacktriangledown}$}
\newcommand{\nosign}{$^{\circ}$}

\begin{document}
\title{Recommending Users:\\Whom to Follow on Federated Social Networks}

\author{Jan Trienes}
\affiliation{%
  \institution{University of Twente}
  \city{Enschede}
  \state{Netherlands}
}
\email{j.trienes@student.utwente.nl}

\author{Andrés Torres Cano}
\affiliation{%
  \institution{University of Twente}
  \city{Enschede}
  \state{Netherlands}
}
\email{a.f.torrescano@student.utwente.nl}

\author{Djoerd Hiemstra}
\affiliation{%
  \institution{University of Twente}
  \city{Enschede}
  \state{Netherlands}
}
\email{d.hiemstra@utwente.nl}

\renewcommand{\shorttitle}{Recommending Users: Whom to Follow on Federated Social Networks}

\begin{abstract}
To foster an active and engaged community, social networks employ recommendation algorithms that filter large amounts of contents and provide a user with personalized views of the network.
Popular social networks such as Facebook and Twitter generate follow recommendations by listing profiles a user may be interested to connect with.
Federated social networks aim to resolve issues associated with the popular social networks --~such as large-scale user-surveillance and the miss-use of user data to manipulate elections~-- by decentralizing authority and promoting privacy.
Due to their recent emergence, recommender systems do not exist for federated social networks, yet.
To make these networks more attractive and promote community building, we investigate how recommendation algorithms can be applied to decentralized social networks.
We present an offline and online evaluation of two recommendation strategies: a collaborative filtering recommender based on BM25 and a topology-based recommender using personalized PageRank.
Our experiments on a large unbiased sample of the federated social network Mastodon shows that collaborative filtering approaches outperform a topology-based approach, whereas both approaches significantly outperform a random recommender.
A subsequent live user experiment on Mastodon using balanced interleaving shows that the collaborative filtering recommender performs on par with the topology-based recommender.
\end{abstract}

\maketitle

\section{Introduction}
Evergrowing concerns about user-privacy, censorship and central authority in popular social media have motivated both the development of federated social networks such as Mastodon and Diaspora~\cite{rochko2018mastodon,Bielenberg:2012:TGD}, as well as research in academia~\cite{Anderson:2009:PSN, Shakimov:2009:PCA}.
These networks aim to promote user control by decentralizing authority and relying on open-source software and open standards.
At the time of this writing, Mastodon has over 1 million users and 3500 instances which demonstrates the increasing acceptance of distributed social networks.
As with traditional social media, one key success factor of such a network is an active and engaged community.

As a community grows, overwhelming amounts of content make it increasingly difficult for a user to find interesting topics and other users to interact with.
For that reason, popular platforms such as Twitter, LinkedIn and Facebook introduce recommender systems that set out to solve a particular recommendation task.
One prominent example is the ``Who to Follow'' service by Twitter~\cite{Gupta:2013:WFS}.
Due to their recent emergence, those recommender systems do not exist for federated social networks, yet.
However, they are needed to make distributed social media attractive to large user groups as well as competitive to centralized networks.
At the same time, recommender systems will contribute to develop, grow and sustain an active community.

To make federated social networks more attractive and feature complete, we implement and evaluate a topology-based user recommender based on personalized PageRank~\cite{Page:1999:PCR}, a commonly used algorithm for link-prediction in social networks.
We compare this method against collaborative filtering based on link intersections~\cite{Hannon:2010:RTU} and a random link predictor baseline~\cite{Liben-Nowell:2003:LPP}.
The experiments are carried out on Mastodon, a federated social network for which user relations do not require reciprocation, and the network forms a directed graph.
We expect that the method and results are transferable to any other federated social network with similar characteristics.

We evaluate the systems in an offline and online scenario. For the offline evaluation, we collect an unbiased sample of the Mastodon user graph.
This sample is created by performing a Metropolis-Hastings Random Walk (MHRW) adapted for directed graphs~\cite{Wang:2011:UGS,Wang:2010:USD}.
The collected data contains about 25\% of the entire userbase of Mastodon.
We then evaluate the recommender systems according to standard performance metrics used in ranked retrieval systems, and deploy the two best performing methods to an online setting.
Both algorithms generate a list of personalized recommendations for 19 Mastodon users participating in the online trial and performance is measured with the balanced interleaving approach~\cite{Joachims:2003:ERP}.

This paper is structured as follows.~\cref{sec:dataset} explains how data are collected for the offline experiments and discusses the recommendation algorithms and their evaluation.
In~\cref{sec:results} we present and discuss experimental results.~\cref{sec:conclusion} concludes this paper and provides directions for future work.

\section{Dataset and Methods}
\label{sec:dataset}

\subsection{Recommendation Algorithms}
\label{sec:recommendation-algorithms}
The user recommendation problem for social networks can be formalized as follows.
Given a graph $G = (V,E)$ where $V$ and $E$ are vertices and edges, we seek to predict an interaction between a user $u \in V$ and $v \in V$ denoted by edge $(u,v)$.
In networks such as Mastodon and Twitter, a user interaction does not require reciprocation.
Thus, the graph is directed.
We consider two broad approaches to generate recommendations: (1) collaborative filtering-based recommendation and (2) topology-based recommendation.

With respect to the collaborative filtering, we use an approach inspired by~\cite{Hannon:2010:RTU}. Each user $u \in V$ is represented by a profile and recommendations are generated based on the similarity of profiles.
We distinguish between the three best performing strategies in~\cite{Hannon:2010:RTU}:
\begin{description}[leftmargin=!,labelwidth=\widthof{$following(u)$}]
	\item[$\mathit{following}(u)$] The set of user ID's $u$ follows
	\item[$\mathit{followers}(u)$] The set of user ID's that follow $u$
	\item[$\mathit{combined}(u)$] The combined set of following and follower ID's
\end{description}
We consider these profiles as documents to be indexed in a general purpose search engine. In order to generate recommendations for a user, the corresponding profile is extracted first.
Afterwards, the retrieval system is queried with the profile and it ranks the indexed documents by their relevance to the query.
Each ID in the user profile is a token of the query. If a  query consists of more than 10,000 tokens, we create a random subset of 10,000 tokens.
Unlike~\citet{Hannon:2010:RTU}, we use BM25 instead of TF-IDF to estimate the relevance score of each document and set parameters to common defaults ($k_1 = 1.2$, $b = 0.75$)~\cite{Manning:2008:IIR:1394399}.
The final recommendation list contains the top-$k$ documents with highest retrieval score.

The collaborative filtering recommendations are compared to topology-based recommendations.
Several methods have been proposed in literature which make use of link-based ranking algorithms such as HITS, PageRank and SALSA\@.
Due to the novelty of generating recommendations for federated social networks, we restrict our experiments to the personalized PageRank algorithm~\cite{Page:1999:PCR} whose efficient computation is well-understood and which is used in the Twitter recommender system~\citep{Gupta:2013:WFS}.
We apply the personalized PageRank for a seed node which is the user we want to generate recommendations for.
After convergence, the list of user recommendations is constructed by taking the top-$k$ nodes with highest PageRank. Following~\cite{Liben-Nowell:2003:LPP}, we set the damping factor $\lambda = 0.85$.

\subsection{Data Collection}
Acquiring the complete graph of a social network is always infeasible due to API limits and time constraints~\cite{Wang:2011:UGS}.
An additional concern arises in a distributed social network.
As data is not stored at a central authority, there is no single API that provides access to all parts of the network.
Instead, data is scattered around different sub-networks.
Both issues are addressed within this section.

To overcome the time constraint, we apply the Metropolis-Hastings Random Walk (MHRW) to acquire an unbiased sample that is still representative of the complete graph.
MHRW is a Markov-Chain Monte Carlo algorithm that can be used to obtain node samples with a uniform probability distribution~\cite{Wang:2011:UGS}.
As the MHRW is only applicable to undirected graphs, we apply a generalization that considers all directed edges as bidirectional edges~\citep{Wang:2010:USD}.
We do not consider graph sampling methods such as Random Walk and Breadth-First Sampling as it has been shown that these methods yield samples biased towards high degree nodes~\cite{Gjoka:2010:WFC}.

Due to the fact that a distributed social network has no central API, one has to query the API of each individual sub-network referred to as \textit{instance}.
In case of Mastodon, there are two public endpoints to acquire incoming and outgoing links: \texttt{/following} and \texttt{/followers}\footnote{The following API URL pattern applies to any Mastodon instance:\\ \texttt{https://<instance>/users/<user>/<endpoint>.json}}.
Whenever the MHRW visits an unexplored node, followers and followings of that node are fetched and stored in a document-oriented database.
This database is also used as a cache: if the random walk transitions to a node which it has already visited, we use the cached result rather than querying the API again.
During the data collection, we apply fair crawling policies. Only instances that allow crawling as defined by the \texttt{robots.txt} are considered.
Furthermore, concurrent requests are throttled such that no more than 10 requests per second are issued (a rate which we believe any web server can sustain).

\subsection{Dataset Statistics}
\label{sec:dataset-statistics}
\cref{tab:dataset-statistics} summarizes the properties of the collected graph. The initial graph ($t_1$) has been crawled from the 16/05/18 until 17/05/18.
The MHRW was executed for 5500 iterations.
During the crawl, 138 instances were disregarded either because of their \texttt{robots.txt} or because they were no longer available.
In order to acquire a newer version of that graph ($t_2$), we visited the same users five days later and recorded new relationships.
The number of visited users in $t_2$ is slightly lower than in $t_1$, as some profiles were deleted or their instances became unavailable.
The updated graph is used as the ground-truth when evaluating our recommender systems.

It can be observed that the Network Average Clustering Coefficient (NCC) and the fraction of nodes in the largest Strongly Connected Component (SCC) is almost equal for the two given graphs.
Furthermore, the graph is mildly disassortative.
It is important to mention that although the total number of nodes found $|V|$ is high (253,000), accounting for about 25\% of the total Mastodon users, the number of visited nodes $|V^*|$ is much smaller (about 3400).
Incoming and outgoing edges are only known for visited nodes.

\begin{table}[t]
\centering
\caption{Statistics of crawled graphs. The initial crawl at $t_1$ and the newer crawl of the same users at $t_2$.}
\label{tab:dataset-statistics}
\begin{tabular}{@{}llllllll@{}}
\toprule
Graph & $|V|$ & $|V^*|$ & $|E|$ & Assort. & Deg. & NCC & SCC \\ \midrule
$t_1$ & 253,822 & 3437 & 754,037 & -0.015 & 5.94 & 0.31 & 0.175 \\
$t_2$ & 255,638 & 3383 & 754,667 & -0.016 & 5.9 & 0.31 & 0.173 \\ \bottomrule
\end{tabular}
\end{table}

\subsection{Evaluation}
\label{sec:evaluation}
The algorithms presented in~\cref{sec:recommendation-algorithms} are evaluated in two phases: an offline evaluation and an online evaluation.
For the offline evaluation we measure precision at rank $k$ (p@k), Mean Average Precision (MAP) and success at rank $k$ (s@k), which are popular metrics for the evaluation of ranked retrieval systems~\cite{Manning:2008:IIR:1394399}.
The newer graph at time $t_2$ serves as the ground-truth, whereas the graph at time $t_1$ can be seen as the training graph.
In information retrieval terms, the generated list of recommendations are the retrieved documents and the list of users a target user follows at time $t_2$ are the relevant documents. Significance is tested using a two-tailed paired t-test. We denote improvements with \upup ($p<0.01$), deteriorations with \downdown ($p<0.01$), and no significance by \nosign.

During the offline evaluation, all systems generate a list of 100 recommendations based on the training graph at time $t_1$.
This list is then compared with the actual links added to the graph in between time $t_1$ and $t_2$ (see~\cref{sec:dataset-statistics}).
In case of the collected dataset, 329 of 3437 visited users started to follow another individual, and thus added a link to the graph.
Only for this set of users, recommendations are generated and evaluated.

The online evaluation is performed as follows.
A recommendation bot is created on the Mastodon instance associated with the institute of the authors\footnote{See \url{https://mastodon.utwente.nl/@Followdon}}.
Afterwards, we ask users to follow this bot if they wish to receive personalized recommendations. For each participant, we generate a static web page consisting of a list of $N$ recommendations with the option to start to follow a suggested user.
A link to this web page is then send to the user and we track the user interactions. A recommendation is considered relevant if the participant starts to follow a suggested user.
The recommendations of two algorithms are presented using balanced-interleaving, which is a relatively inexpensive evaluation method for online experiments compared to conventional A/B testing.
We refer the reader to~\cite{Joachims:2003:ERP} for a thorough discussion of this evaluation method.

One complication arises in the online evaluation. As an up-to-date graph is unavailable at recommendation time, such a graph has to be created.
For this, we explore the vicinity of a recommendation target $u \in V$ by applying an egocentric random walk for a fixed amount of iterations.
This strategy resembles the ``circle-of-trust'' used in the Twitter recommender system~\cite{Gupta:2013:WFS}.
The random walk is performed as follows. At each iteration, the algorithm either transitions to a random neighbor of the current user with probability $\gamma$, or jumps back to $u$ with probability $1 - \gamma$. In our experiments, we execute the random walk for 200 iterations and set $\gamma = 0.8$.
Here, we do not claim that this is the most efficient way of generating recommendations in an online setup. It is merely a way to deal with incomplete data in federated social networks.

\section{Results}
\label{sec:results}

\subsection{Offline Evaluation}
\begin{figure}[t]
\centering
\includegraphics[width=.8\columnwidth]{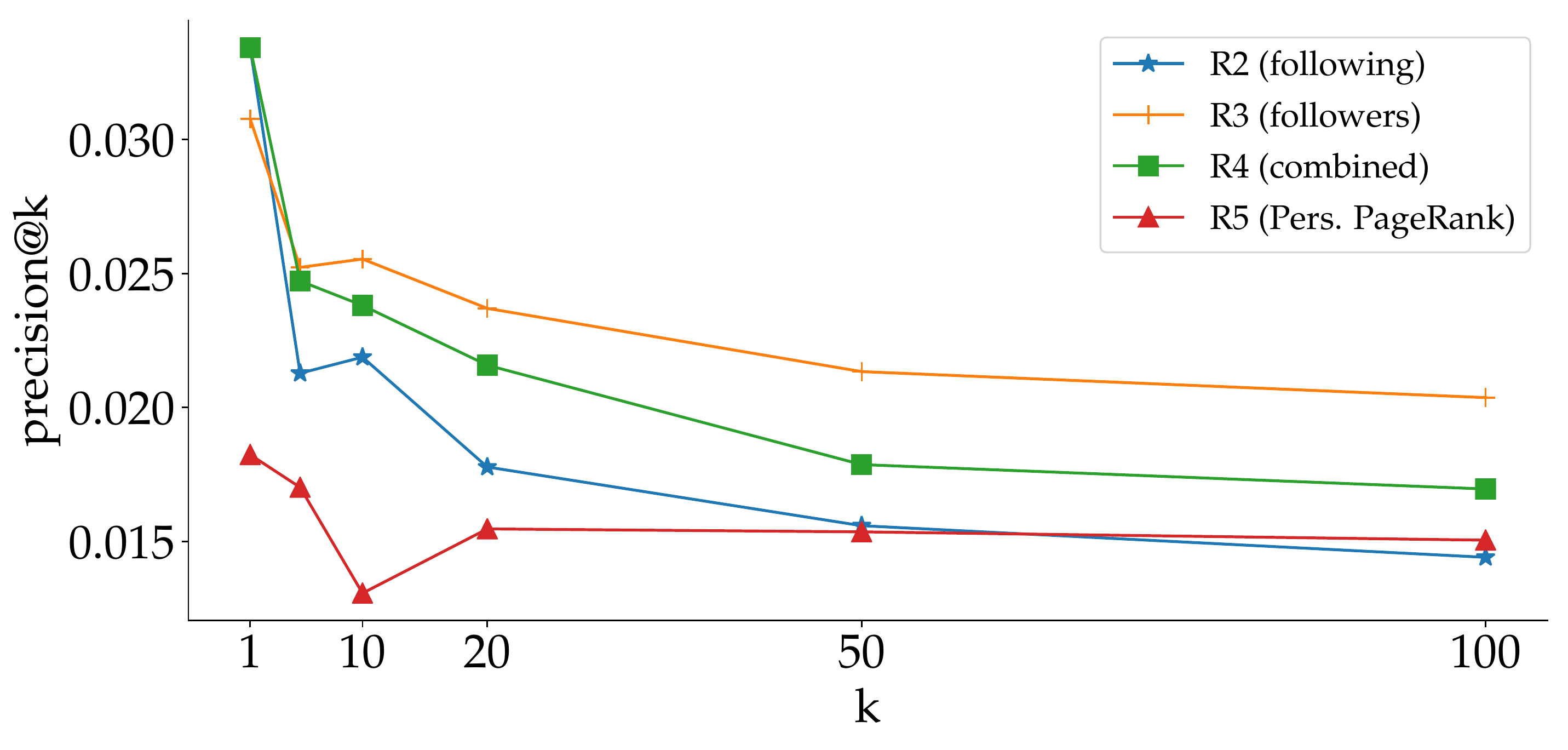}
\caption{Precision for different recommendation list lengths ($k$) in offline evaluation.}
\label{fig:precision-at-k}
\end{figure}

The collaborative filtering approach shows a consistently higher performance than a topology-based system using PageRank (see~\cref{tab:results-offline}).
With respect to the success at rank $k$ metric, profile-based approaches (R2--R4) have up to two times higher retrieval scores than PageRank (R5).
The individual profiling strategies perform all rather similarly, which aligns with the findings in~\cite{Hannon:2010:RTU}.
Also, a baseline system (R1) which generates recommendations by selecting 100 random users from the network topology is outperformed by a large margin.
In \cref{fig:precision-at-k}, it can be observed that shorter recommendation lists have a higher precision for all recommendation strategies.
Precision at rank $k$ remains stable starting from a list length of $k = 50$ items.
This suggests that shorter lists are to be preferred in an online scenario.

\begin{table}[t]
\centering
\caption{Experimental results of offline evaluation. Significance for model in line $i > 1$ is tested against line $i - 1$.}
\label{tab:results-offline}
\begin{tabular}{@{}clllll@{}}
\toprule
\textbf{ID} & \textbf{System}     & \textbf{MAP} & \textbf{s@1} & \textbf{s@5} & \textbf{s@10} \\ \midrule
R1          & Random              & 0.001       & 0.000            & 0.000            & 0.055         \\
R2          & Profile (following) & \textbf{0.019}\upup       & \textbf{0.033}\upup        & 0.085\upup        & 0.152\upup         \\
R3          & Profile (followers) & \textbf{0.019}\nosign       & 0.030\nosign        & 0.100\nosign        & 0.167\nosign         \\
R4          & Profile (combined)  & 0.018\nosign       & \textbf{0.033}\nosign        & \textbf{0.106}\nosign        & \textbf{0.173}\nosign         \\
R5          & Pers.\ PageRank     & 0.014\nosign       & 0.018\nosign        & 0.061\downdown        & 0.082\downdown         \\ \bottomrule
\end{tabular}
\end{table}

It is important to mention that the list of possible suggestions from the profile-based recommender is smaller than the list from the PageRank recommender, which complicates the discussion.
Only visited nodes (see~\cref{sec:dataset-statistics}) have been indexed in the document retrieval system.
This significantly reduces the pool size of possible users ($\approx$3k).
In contrast, the PageRank recommender can suggest any user in the topology ($\approx$255k).
One could overcome this issue as follows.
Each user, regardless of whether or not it has been visited during the data collection, could be added to the search index.
Then, incoming relationships can be inferred by inspecting the outgoing links of visited users.
By adding these relations as \textit{followers} to the documents of unexplored nodes, the \textit{following} strategy of the collaborative filtering can be applied.
However, as no outgoing links are known for unexplored users, the \textit{following} and \textit{combined} strategies are not fully applicable.
Due to time constraints, we did not further investigate this issue.

Furthermore, it is worth to note that the chosen window of five days between $t_1$ and $t_2$ might not have been long enough to capture sufficient user activity.
In between the training snapshot at $t_1$ and the testing snapshot at $t_2$, six new connections were added to each user on average.
This gives rise to an interesting trade-off.
For a longer time span, one can capture larger amounts of activity within the network.
Intuitively, more links will be added as users start to follow other users.
However, the farther two snapshots are apart, the larger is the risk that the network deviates too much from the original structure.
Users might stop following other users or profiles could be deleted.
More severely, entire instances could become unavailable due to a temporary downtime, or they could even be discontinued.
This is a unique concern related to the distributed nature of federated social networks.

Finally, we want to motivate which recommendation systems are evaluated in the online trial based on the results presented above.
From~\cref{tab:results-offline} it can be observed that the profile-based recommendation strategies perform rather similarly.
However, the combined strategy (R4) performs best with respect to the success at rank 10 metric, which one seeks to maximize in an online system where 10 recommendations are presented to the user.
Therefore, we pick R4 as the first recommendation system.
Although the personalized PageRank recommender (R5) has a lower performance than the other profiling strategies, we expect that it produces valuable recommendations which are significantly different from the profile-based strategies.
This is due to the fact that it considers the network topology when generating recommendations.
Therefore, we apply the balanced-interleaving evaluation to systems R4 and R5.

\subsection{Online Evaluation}
The online evaluation shows that neither the profile-based nor the topology-based system is superior (see~\cref{tab:summary-online-evaluation}). Nineteen users participated in our online study. On average, they started to follow 1.8 users from our recommendations. For 5 users the profile-based approach performed best. For another 5 users, the topology-based approach performed best. For the remaining 9 users both system performed equally well, or no recommendation was followed.
The fact that valuable recommendations were generated that resulted in new followings shows that the two systems can be useful in practice.
However, a larger group of participants is required to draw final conclusions on the recommender system performance.

\begin{table}[t]
\centering
\caption{Summary of online evaluation.}
\label{tab:summary-online-evaluation}
\begin{tabular}{@{}p{.7\columnwidth}c@{}}
\toprule
\textbf{Characteristic}  & \textbf{Value} \\ \midrule
Number of participants   & 19 \\
Profile-based recommender (R4) superior & 5\\
PageRank recommender (R5) superior & 5\\
Draw & 2 \\
No user interaction & 7 \\
 \bottomrule
\end{tabular}
\end{table}

\subsection{Practical Considerations}
The generation of online recommendations turned out to be costly because the complete network data is not available.
In contrast to centralized social media, federated social networks do not have a single authority which stores data about the entire network graph.
The proposed method of crawling the vicinity of a target user at recommendation time (see~\cref{sec:evaluation}) comes with a high overhead in network traffic and is not suitable for real-time systems that have to support large amounts of users.
In addition to that, the method is sensitive to the size of the vicinity.
We expect that a larger number of iterations yields a better picture of a user's vicinity, which in turn increases the quality of recommendations. However, an exploration of different parameter settings has been out of scope of this study.
The data collection issue is even more severe in the offline evaluation which requires large and representative samples of the entire network.

To reduce the overhead associated with crawling in an online setting, one might attempt to gradually construct a cached representation of the entire network graph.
Whenever a recommendation is generated for a user, the vicinity is added to that graph. On subsequent recommendations, one might reuse parts of this network to avoid additional crawling.
This approach has two important issues that have to be considered. First, one has to address the question when parts of the network are considered to be out of date (i.e., when the cache expires). Second, and more importantly, such an approach seems to be in conflict with the intentions behind decentralization.
By constructing a database that aims to capture the entire network graph, one starts to centralize the data of a federated social network.\looseness=-1

\section{Conclusion}\label{sec:conclusion}
User recommendation algorithms commonly applied to centralized social media can be applied to incomplete data from federated social networks with the goal of developing an engaged community.
We showed that collaborative filtering-based recommenders outperform a topology-based recommender on a large unbiased sample of the federated social network Mastodon. The two recommenders outperform a random recommender by a large margin.
A subsequent live user experiment on Mastodon using balanced interleaving shows that the two recommender approaches perform on par.
Acquiring a sufficiently large snapshot of the network topology for offline recommendation proofed to be difficult and costly. Keeping the snapshot up-to-date needs constant re-sampling. Online recommendation was done by sampling the graph neighborhood for the current user.

There are several directions for future work. First, studying the extent to which incomplete data impacts the recommender performance may derive methods that are tailored towards federated social networks which operate with limited amounts of data.
Second, user recommendation algorithms in popular social media increasingly utilize user context information such as location data and interests.
It remains unclear how such data can be effectively acquired and utilized in federated social networks while preserving privacy.
Third, BM25 might not be the best ranking function for the presented  recommender approach, and it should be compared to functions that also use popularity-based scoring.
Finally, one may investigate how decentralized communication protocols such as ActivityPub can be extended to support community building algorithms while maintaining the notion of decentralized network data.\looseness=-1


\bibliographystyle{ACM-Reference-Format}
\bibliography{bibliography}

\end{document}